\documentclass[aps,twocolumn,superscriptaddress,a4paper]{revtex4}
\usepackage{graphics}
\usepackage{amsmath}
\begin{document}

\title{An integral equation approach to effective interactions between
polymers in solution}

\author{V. Krakoviack} \email{Vincent.Krakoviack@ens-lyon.fr}
\affiliation{Laboratoire de Chimie, \'Ecole Normale Sup\'erieure de
Lyon, 46 All\'ee d'Italie, 69364 Lyon cedex 07, France.}

\author{B. Rotenberg} \affiliation{Department of Chemistry, University
of Cambridge, Lensfield Road, Cambridge CB2 1EW, United Kingdom.}
\affiliation{D\'epartement de Chimie, \'Ecole Normale Sup\'erieure, 24
rue Lhomond, 75005 Paris, France.}

\author{J.-P. Hansen} \affiliation{Department of Chemistry, University
of Cambridge, Lensfield Road, Cambridge CB2 1EW, United Kingdom.}

\date{\today}

\begin{abstract}
We use the thread model for linear chains of interacting monomers, and
the ``polymer reference interaction site model'' (PRISM) formalism to
determine the monomer-monomer pair correlation function $h_{mm}(r)$
for dilute and semi-dilute polymer solutions, over a range of
temperatures from very high (where the chains behave as self-avoiding
walks) to below the $\theta$ temperature, where phase separation sets
in. An inversion procedure, based on the HNC integral equation, is
used to extract the effective pair potential between ``average''
monomers on different chains. An accurate relation between
$h_{mm}(r)$, $h_{cc}(r)$ [the pair correlation function between the
polymer centers of mass (c.m.)], and the intramolecular form factors
is then used to determine $h_{cc}(r)$, and subsequently extract the
effective c.m.-c.m. pair potential $v_{cc}(r)$ by a similar inversion
procedure. $v_{cc}(r)$ depends on temperature and polymer
concentration, and the predicted variations are in reasonable
agreement with recent simulation data, except at very high
temperatures, and below the $\theta$ temperature.
\end{abstract}


\maketitle

\section{Introduction}

The Statistical Mechanics of dilute and semi-dilute solutions of
interacting polymers may be greatly simplified by adopting a
coarse-graining strategy for this complex many-body system, whereby
the initial monomer or segment-level description is reduced to a
``soft-colloid'' representation involving only the centers of mass
(c.m.) of the polymer coils.  This is formally achieved by tracing out
the individual monomer degrees of freedom for fixed positions of the
c.m.'s of the polymers. This reduction procedure results in effective
interactions between the c.m.'s which are state dependent, since they
are in fact \textit{free} energies of the ``fluid'' of monomers, for
each configuration of the c.m.'s of the $N$ interacting coils.  In the
low concentration limit, this free energy reduces to a pairwise
additive sum of effective potentials of mean force acting between the
c.m.'s of overlapping polymers, which depend only on temperature. Such
pair potentials were first considered by Flory and Krigbaum
\cite{flokri50jcp}, and later put on a firmer basis by Grosberg
\textit{et al.} \cite{grokhakho82mcrc}, who showed that the effective
pair potential between the c.m.'s of linear self-avoiding walk (SAW)
polymers remains finite at full overlap in the scaling limit (where
the number of segments $L\rightarrow\infty$), and of the order of a
few $k_{B}T$, reflecting the purely entropic origin of the effective
interaction in the athermal SAW model.  Quantitative estimates of the
effective c.m. pair potential between two isolated SAW polymers were
obtained by Monte Carlo (MC) simulations of on- and off-lattice models
\cite{olalanpel80mm,dauhal94mm,loubolhanmei00prl,bollouhanmei01jcp%
,bollou02mm} and by renormalization group calculations
\cite{kruschbau89jpf}.  These studies show that the zero concentration
pair potential is purely repulsive and reasonably well represented by
a Gaussian of amplitude $\simeq 2 k_{B}T$ and width equal to the
radius of gyration $R_{g}$.

The SAW model describes polymers in good solvent. Poor solvent
conditions may be modeled by adding a finite attractive interaction
between monomers and lowering the temperature. MC simulations
\cite{dauhal94mm,krahanlou03pre} show that the resulting effective
c.m. pair potential becomes less repulsive as the temperature $T$ is
lowered, and can become attractive in the vicinity of the $\theta$
temperature.

When the polymer concentration is increased, three- and more-body
effective interactions between the polymer c.m.'s come into
play. These can also in principle be calculated by MC simulations of
clusters of three or more polymers \cite{bollouhan00pre}, but this
systematic procedure becomes rapidly untractable, since the effective
interactions depend on an increasing number of variables.

An alternative approach is to restrict the effective interactions
between polymer c.m.'s at finite concentrations to a sum of pair
potentials, the latter being determined by an inversion procedure of
the c.m. pair distribution function based on integral equations for
the pair structure of classical fluids
\cite{loubolhanmei00prl,bollouhanmei01jcp,bollou02mm,krahanlou03pre}.
The price to pay is that the resulting pair potentials now depend on
polymer concentration, in addition to temperature. The variation with
concentration is moderate, but significant, in the case of the SAW
model for polymers in good solvent
\cite{loubolhanmei00prl,bollouhanmei01jcp,bollou02mm}, and becomes
very strong for polymers in poor solvent \cite{krahanlou03pre}.

Since the effective pair potentials are obtained by inverting
c.m. pair correlation data from MC simulations, this strong variation
with polymer concentration means that full monomer-level simulations
must be carried out for each thermodynamic state of the polymer
solution, thus partly defeating the purpose of the coarse-graining
strategy.

It is hence important to explore numerically less demanding
theoretical methods for the determination of the required polymer pair
correlation functions. One obvious candidate is the polymer reference
interaction site model (PRISM) theory, which has proved very
successful to determine the pair structure of interacting polymer
solutions and melts \cite{prism}. PRISM grew out of the RISM integral
equation of Chandler and Andersen \cite{chaand72jcp} for the pair
structure of simple molecular fluids. The theory includes
intramolecular connectivity via the macromolecular form factor, and
provides accurate monomer-monomer pair distribution functions,
$g_{mm}(r)$. The c.m.-c.m. pair distribution function, $g_{cc}(r)$,
which needs to be inverted to extract the effective c.m.-c.m. pair
potential, $v_{cc}(r)$, can be derived from $g_{mm}(r)$ by a simple,
accurate relation which was established recently within the PRISM
framework \cite{krahanlou02el}.

This strategy, which is highly economical on computational resources,
is explicitly carried out in this paper, on the basis of the Gaussian
thread model for interacting polymers
\cite{schcur88mm,chasch98jcp,chasch98mm,cha93pre}. PRISM theory, in
conjunction with an approximate closure relation, allows the
monomer-monomer pair distribution function $g_{mm}(r)$ to be
calculated analytically or numerically depending on the expression for
the polymer form factor. The c.m. pair distribution function is then
extracted from $g_{mm}(r)$ using the relation of
Ref.~\cite{krahanlou02el} together with appropriate form factors, and
$v_{cc}(r)$ is finally obtained by hypernetted-chain (HNC) inversion
\cite{loubolhanmei00prl,bollouhanmei01jcp,bollou02mm,krahanlou03pre}. The
variation of the effective pair potential with temperature and polymer
concentration is then examined in detail, and compared to the
``exact'' results extracted from MC simulations
\cite{loubolhanmei00prl,bollouhanmei01jcp,bollou02mm,krahanlou03pre}.

\section{Polymer model and PRISM theory}\label{prism.sec}

We consider solutions of $N$ linear polymers made up of $L$ freely
jointed monomers or segments of diameter $d$ and spacing $\sigma$.
Monomers belonging to the same or different chains are not allowed to
overlap, i.e. they experience an infinite mutual repulsion for
monomer-monomer distances $r<d$, and, for $r>d$, they attract each
other via a Yukawa interaction,
\begin{equation}\label{yukawa.eq}
v(r)=-\epsilon\frac{a}{r}e^{-r/a},
\end{equation}
where $\epsilon>0$ is the energy scale and $a$ the range of the
monomer-monomer attraction \cite{chasch98jcp,chasch98mm}. The relevant
thermodynamic variables are the temperature $T$ [as usual, we define
$\beta=(k_BT)^{-1}$] and the polymer concentration $\rho$ (number of
polymer coils per unit volume); the corresponding monomer density is
$\rho_{m}=\rho L$. As usual, the spatial extension of the chain will
be characterized by its radius of gyration $R_g$ and the polymer
overlap concentration, at which different chains start to overlap, is
defined as $\rho^*=(\frac43\pi R_g^{3})^{-1}$.  In this model of
polymer solutions, the attractive interaction (\ref{yukawa.eq})
represents solvent mediated monomer-monomer interactions in an
effective way. Then, with respect to intermolecular interactions,
large $T$ values correspond to good solvent conditions, while lower
$T$ values mimic poor solvent conditions.

PRISM theory assumes all monomers to be equivalent regardless of their
position along the polymer chain. Its aim is to calculate the
site-averaged pair distribution function $g_{mm}(r)$ of monomers on
different chains. To this end one introduces the pair correlation
function, $h_{mm}(r)=g_{mm}(r)-1$, the direct correlation function
$c_{mm}(r)$, and their Fourier transforms, respectively
$\widehat{h}_{mm}(q)$ and $\widehat{c}_{mm}(q)$, which are related by
the following generalization of the Ornstein-Zernike (OZ) equation
\cite{prism},
\begin{equation}\label{prismoz.eq}
\widehat{h}_{mm}(q) = \widehat{\omega}_{mm}(q) \widehat{c}_{mm}(q)
[\widehat{\omega}_{mm}(q) + \rho_{m} \widehat{h}_{mm}(q)],
\end{equation}
where $\widehat{\omega}_{mm}(q)$ is the form factor of the polymer
chains. In the following, it will be convenient to introduce the
auxiliary quantities
$\widehat{\Omega}_{mm}(q)=\widehat{\omega}_{mm}(q)/L$ [hence,
$\widehat\Omega_{mm}(q=0)=1$] and
$\widehat{C}_{mm}(q)=L^{2}\widehat{c}_{mm}(q)$, such that
Eq.~(\ref{prismoz.eq}) may be rewritten as
\begin{equation}\label{prismozthread.eq}
\widehat{h}_{mm}(q) = \widehat{\Omega}_{mm}(q) \widehat{C}_{mm}(q)
[\widehat{\Omega}_{mm}(q) + \rho \widehat{h}_{mm}(q)].
\end{equation}
The OZ equation, Eq.~(\ref{prismozthread.eq}), must be supplemented by
a closure relation for $\widehat{C}_{mm}(q)$, as well as by a
prescription for $\widehat{\Omega}_{mm}(q)$. While self-consistent
schemes exist to compute the latter as a function of temperature and
density, they are quite complex to implement \cite{prism}. In the
present work, we will make the simplifying assumption that the
polymers obey Gaussian statistics. This approximation, which is in
principle only valid in the melt or close to the $\theta$ temperature,
is mainly needed for technical reasons, due to the lack of analytical
results for non-Gaussian polymers. Thus, the form factor
$\widehat{\Omega}_{mm}(q)$ is given by the well-known Debye expression
for ideal chains in the continuum limit, namely \cite{doiedwards}
\begin{equation}\label{debye.eq}
\widehat{\Omega}_{mm}(q)=\frac{2}{x^{4}}\left[e^{-x^2}-1+x^{2}\right],\
x=qR_g.
\end{equation}
Replacing this exact form by its simple Pad\'e approximant,
\begin{equation}\label{omegaapp.eq}
\widehat{\Omega}_{mm}(q)=\frac{1}{1+x^2/2},\ x=qR_g,
\end{equation}
allows analytic results to be obtained within the thread limit
\cite{schcur88mm,chasch98jcp,chasch98mm}, but we will see in the
following that it can lead to significant problems in the present
context.

In this work, we are interested in a mesoscopic level of description
of polymer systems. For simplicity, it is thus useful to get rid of
the microscopic details of the polymer model. A way of doing this in
the PRISM framework is by taking the so-called thread limit. In the
case of excluded volume interactions only, it is defined by letting
$d$ and $\sigma$ go to zero and $L$ to infinity, keeping the ratio
$\Gamma=\sigma/d$, $R_{g}$ and $\rho/\rho^*=\frac43\pi\rho R_g^3$
constant
\cite{schcur88mm,chasch98jcp,chasch98mm,fucmul99pre,fucsch01pre}. In
Refs.~\cite{chasch98jcp,chasch98mm}, the attractive part of the
monomer-monomer interaction potential was left untouched when taking
the thread limit. However, for simple polymers, it turns out that the
range of this attractive tail scales with the monomer size. Thus for
this work it seemed physically appropriate to supplement the previous
limiting process with the limit of a vanishing range of the attractive
potential, namely $a\to0$, keeping the ratio $\Delta=a/\sigma$
constant. This can be interpreted as a special limit of the theory of
Refs.~\cite{chasch98jcp,chasch98mm}, as shown in Appendix
\ref{thread.app}.

\begin{figure}
\includegraphics{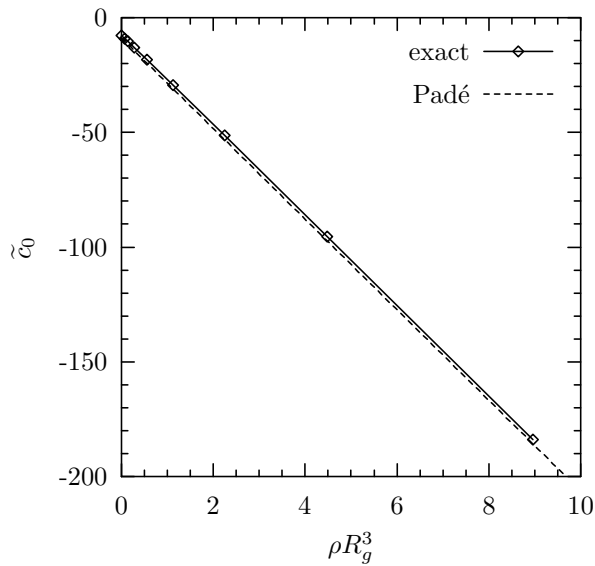}
\caption{\label{czero.fig} Density dependence of the direct
correlation function $\widetilde{c}_{0}$, in the presence of excluded
volume interactions only. For the exact form factor (\ref{debye.eq}),
$\widetilde{c}_{0}$ has to be numerically determined, while for the
approximation (\ref{omegaapp.eq}), the analytic result (\ref{czero.eq})
is plotted.}
\end{figure}

To escape the trivial case of non-interacting polymers, the thread
limit requires a proper formulation of the closures of the PRISM-OZ
equation (\ref{prismozthread.eq}) as well. In this work, we have
applied the so-called RMPY/HTA closure, which has been found to be the
most satisfactory in earlier work
\cite{chasch98jcp,chasch98mm,schyet93jcp}.  The details are given in
Appendix \ref{thread.app} and we only quote here the equations
relevant to our calculation.  The monomer-monomer correlations are
thus given by
\begin{equation}\label{h.eq}
\widehat{h}_{mm}(q) = \frac{\widetilde{c} R_g^3
\widehat{\Omega}_{mm}(q)^{2}}{1-\rho\widetilde{c}R_g^3
\widehat{\Omega}_{mm}(q)},
\end{equation}
with
\begin{equation}\label{c.eq}
\widetilde{c}= \widetilde{c}_0  \left(1-\frac{\Theta}{T}\right),
\end{equation}
where $\Theta$ is the $\theta$ temperature of the polymer model.
$\widetilde{c}_{0}$, which depends on $\rho R_g^3$ only, is obtained by
imposing that, in the presence of excluded volume interactions only
(i.e. for infinite temperature), $h_{mm}$ obeys the hard core
condition, $h_{mm}(r=0)=-1$. It is plotted for both chain form
factors, Eq.~(\ref{debye.eq}) and (\ref{omegaapp.eq}), in
Fig.~\ref{czero.fig}.

With the above monomer pair correlation functions, effective pair
potentials between polymer chains may now be determined. We first
consider in the next Section the situation where the interaction site
for the effective interaction between polymer coils is chosen to be a
monomer; subsequently effective interactions between polymer c.m.'s
are discussed in Sec.~\ref{vcc.sec}.

\section{Effective interaction between monomers}\label{vmm.sec}

Starting from the original polymer system, the choice of the effective
coordinates used to specify the position of each polymer coil, and
relative to which the coarse-graining procedure is performed, is
largely arbitrary. Most of the time, the center of mass is considered,
but it has sometimes been suggested to use specific monomers, like the
midpoint or the endpoints, as in
Refs.~\cite{jusdzulikferlow01jpcm,bollouhanmei01jcp,bollou02mm}. This
is a very natural choice for star polymers as well, where the star
center provides an obvious reference point \cite{lik01pr}. In this
Section, we thus attempt to build such a monomer-monomer (or
site-site) effective interaction.

In the present PRISM-based approach, by construction, no distinction
is made between the different monomers on a chain.  However, one can
build an effective monomer-monomer interaction based on the concept of
an average monomer \cite{sch02pre}. The idea goes as follows. On each
chain, a random monomer is tagged, and a coarse-grained
polymer-polymer interaction is computed by tracing out, for fixed
(spatial) positions of the tagged monomers, all the other monomer
degrees of freedom. Averaging this free energy over the random
locations of the tagged monomers on their chains then leads to a
quantity which can be given the meaning of an effective interaction
between average monomers.

In practice, an effective pair potential $v_{aa}(r)$ may be extracted
by an inversion procedure, based on integral equations for the pair
structure of simple fluids \cite{simpleliquids}. For this purpose, we
consider a fluid of average monomers, of density $\rho$ (each polymer
chain is originally associated with a single tagged interaction site),
the pair correlations of which are characterized by $h_{aa}(r) \equiv
h_{mm}(r)$, where $h_{mm}(r)$ is the monomer-averaged correlation
function determined in the previous section \cite{order}. A direct
correlation function $c_{aa}(r)$ may be extracted from $h_{aa}(r)$ via
the OZ relation for simple (atomic) fluids which in Fourier space
reads
\begin{equation}\label{oz.eq}
\widehat{h}_{aa}(q) = \widehat{c}_{aa}(q) [ 1 + \rho
\widehat{h}_{aa}(q) ].
\end{equation}
The effective interaction potential may finally be determined using
one of the standard closures for simple fluids. Since the effective
interactions are expected to be relatively weak (in particular, they
contain no hard core repulsion), we follow earlier work
\cite{loubolhanmei00prl,bollouhanmei01jcp,bollou02mm,krahanlou03pre},
and adopt the very accurate HNC closure, which leads to the following
expression for $v_{aa}(r)$:
\begin{equation}\label{hnc.eq}
\beta v_{aa}(r) = h_{aa}(r)-c_{aa}(r)-\ln[1+h_{aa}(r)].
\end{equation}
With this closure, a fully analytic expression for $\beta v_{aa}(r)$
can be obtained if the approximate expression (\ref{omegaapp.eq}) for
the chain form factor is used. Details are given in Appendix
\ref{anavaa}.

\begin{figure*}
\includegraphics{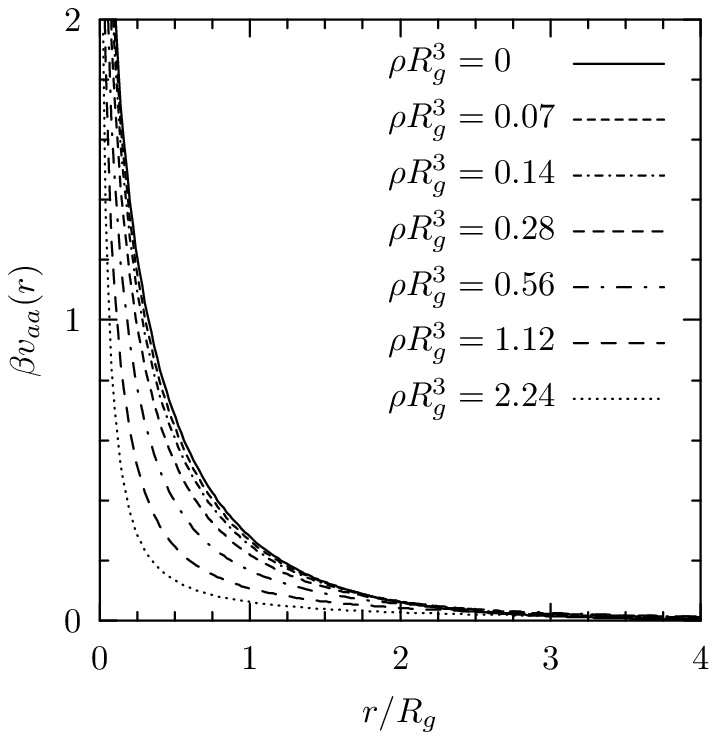}
\hspace{0.5cm}
\includegraphics{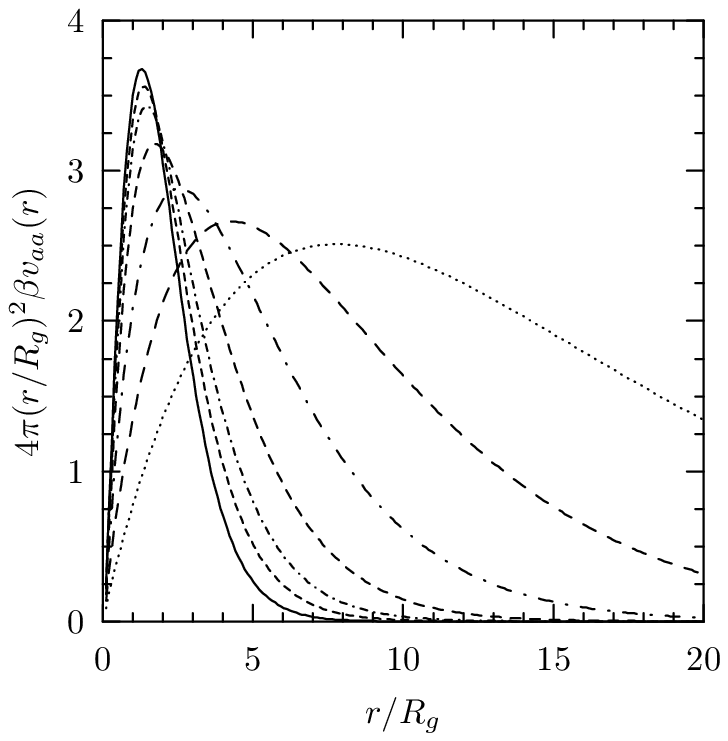}
\caption{\label{vaa.fig} Effective pair potentials between average
monomers (left-hand frame), obtained with the approximate polymer form
factor (\ref{omegaapp.eq}), at various reduced densities $\rho R_g^3$,
in the presence of excluded volume interactions only; the right-hand
frame shows the potentials multiplied by $4\pi(r/R_g)^2$.}
\end{figure*}

We first consider the case of excluded volume interactions only,
i.e. $\widetilde{c}=\widetilde{c}_0$.  The analytic $\beta v_{aa}(r)$
is plotted in Fig.~\ref{vaa.fig} for various polymer densities. It
diverges at zero separation and, when the density increases, this
divergence manifests itself at shorter distances and becomes
steeper. At the same time, the repulsive tail of the potential becomes
longer-ranged, as can be clearly seen on the plot of
$4\pi(r/R_g)^2\beta v_{aa}(r)$, a quantity playing a crucial role for
the calculation of the thermodynamic properties of fluid systems
characterized by similar ``soft'' interaction potentials
\cite{lanlikwatlow00jpcm}.  These results are in good qualitative
agreement with the simulation data of Bolhuis and Louis for midpoint
and endpoint effective pair potentials \cite{bollou02mm}.

Recently, much attention has been devoted to the development of
analytic expressions for effective pair potentials between the centers
of star polymers in solution \cite{lik01pr}. Since a linear polymer,
seen from one of its endpoints or from its midpoint, can be identified
with a one- or two-arm star, respectively, it is interesting to
compare those calculations with the present theory.

Since, according to Eq.~(\ref{hzero.eq}), $1+h_{mm}(r)$ vanishes
linearly for $r\to0$, $\beta v_{aa}(r)$ diverges like
$-\ln(r/R_g)$. Such a weak, logarithmic divergence is present in the
star polymer models, where it has been derived using scaling
arguments, but the prefactor, which should be given by the two-arm
result in the limit of very long chains where end effects are expected
to be negligible, appears overestimated by the present theory compared
to scaling arguments (which predict a prefactor equal to $5
f^{3/2}/18$ for a star with $f$ arms) and renormalization group
results \cite{mulbinsch00mm}.  This quantitative disagreement is not
surprising, since the present calculation relies on the use of
Gaussian form factors in good solvent conditions.  However, a similar
calculation done with an approximate form factor reproducing the
correct asymptotic behavior of $\widehat{\Omega}_{mm}(q)$ in good
solvent still does not provide the correct value and this failure was
in fact argued to be inherent to the thread PRISM approach
\cite{fucmul99pre}.

For large distances, $\beta v_{aa}(r)$ asymptotically takes a Yukawa
form \cite{zeroexp},
\begin{equation}
\beta v_{aa}(r) \propto \frac{e^{-r/\xi_{+}}}{r/R_g}.
\end{equation}
Such a long range behavior was postulated for star polymers, on the
basis of the blob picture of the star inner structure. However,
because the blob picture breaks down for small $f$, it was advocated
that this asymptotic behavior should not be valid for small
functionalities, $f\lesssim10$, and should be replaced by a Gaussian
tail, by analogy with the behavior of the effective pair potential
between chain centers of mass \cite{jusdzulikferlow01jpcm}. The
present calculation does not appear to support this choice.

Finally, and maybe more importantly, the present calculation reveals
that $\beta v_{aa}(r)$ changes significantly with density. For
instance, the range of the asymptotic Yukawa tail more than doubles
between $\rho/\rho^*=0$ and $\rho/\rho^*=1$. If such a behavior also
applies to star polymers, it could have a major impact on the
calculation of the thermodynamic properties of these systems, in
particular on their phase diagrams.

\begin{figure}
\includegraphics{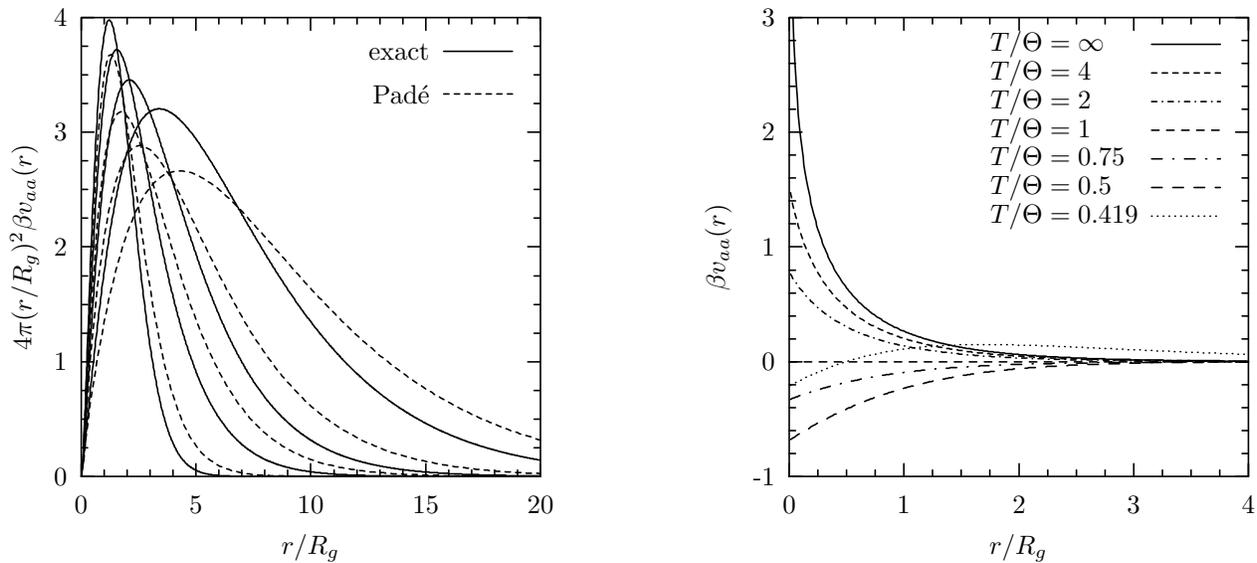}
\caption{\label{vaacomp.fig} Comparison of the effective pair
potentials between average monomers [multiplied by $4\pi(r/R_g)^2$]
obtained with the exact and approximate polymer form factors,
(\ref{debye.eq}) and (\ref{omegaapp.eq}) respectively, at various
densities, in the presence of excluded volume interactions only. From
left to right: $\rho R_g^3=0,\ 0.28,\ 0.56,\ 1.12$.}
\end{figure}

In Fig.~\ref{vaacomp.fig}, the numerically determined $\beta
v_{aa}(r)$ for the exact Debye form factor (\ref{debye.eq}) is
compared to the analytic result obtained with the Pad\'e approximant
(\ref{omegaapp.eq}), at various polymer densities. A thorough analysis
shows that the differences between both are only quantitative. Both
share the same $-\ln(r/R_g)$ divergence for small separations, and
both display an asymptotic Yukawa tail, which is of shorter range when
(\ref{debye.eq}) is used.

\begin{figure}
\includegraphics{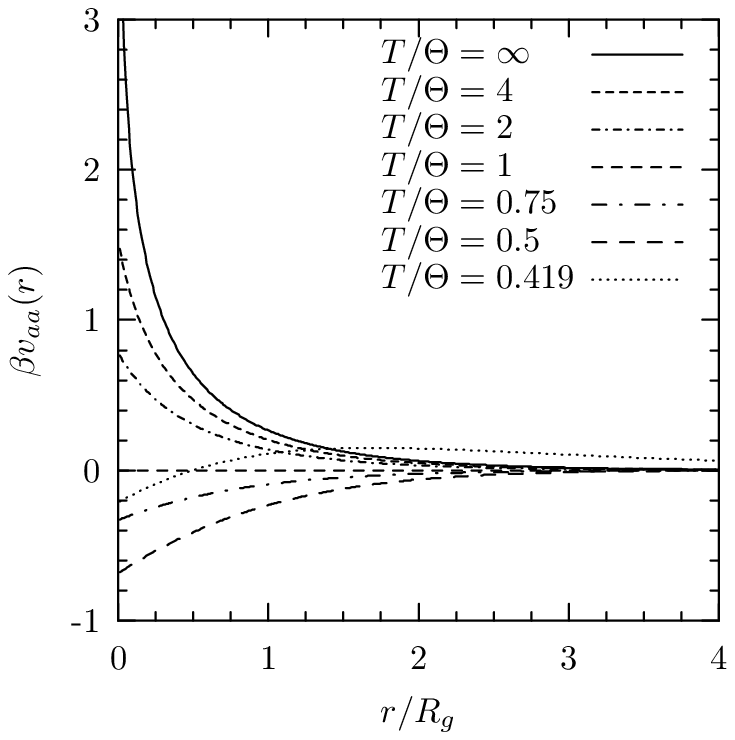}
\caption{\label{vaatemp.fig} Effective pair potentials between average
monomers, obtained with the approximate polymer form factor
(\ref{omegaapp.eq}), at various temperatures, for the reduced density
$\rho R_g^3=0.07$.}
\end{figure}

We now turn to the effect of the attractive monomer-monomer
interaction with changing temperature. In Fig.~\ref{vaatemp.fig}, the
analytic $\beta v_{aa}(r)$ calculated with the approximate chain form
factor (\ref{omegaapp.eq}) is plotted at density $\rho R_g^3=0.07$ for
various values of the temperature. Similar results are obtained at all
densities, and with the exact Debye form factor as well. The effective
potential is seen to become less repulsive when the temperature is
decreased from $\infty$ to $\Theta$, it vanishes identically at
$\Theta$ and grows more and more attractive if the temperature is
further decreased. This trend reverts when the ``spinodal'' line is
approached, leading to non-monotonous, mostly repulsive effective
potentials of increasing range.

But a word of caution is needed here. Indeed, as shown in Appendix
\ref{thread.app}, at all finite temperatures, the hard core condition
$h_{mm}(r=0)=-1$ is no longer obeyed. Thus, the effective pair
potential between average monomers does not diverge at full overlap as
it should, if this condition was consistently enforced at all state
points. This might not be so important, since, as it has already been
mentioned, the product $4\pi(r/R_g)^2\beta v_{aa}(r)$ is in general
more relevant than the potential itself, and in this quantity the
short range details of the potential are strongly reduced by the
vanishing prefactor. Another source of concern is that it is not known
to what extent the results obtained below $\Theta$ when approaching
the ``spinodal'' line are physically meaningful, since the
corresponding remnant of a binodal line, which would properly identify
the domain of existence of the low density phase, is not known in the
present theory \cite{binodal}. For these reasons, the present results
are only given for reference and would need to be validated, both by
implementing alternative integral equation schemes or by investigating
$\beta v_{aa}(r)$ as a function of temperature by molecular
simulation.

To summarize, we have shown in this Section that integral equation
theories for macromolecular systems, like the present thread PRISM
theory, could be useful tools to compute monomer-based effective
interaction potentials between polymers. At least in the case of
excluded volume interactions only, they lead to very reasonable
predictions for the effective interactions between average monomers on
polymer chains, compared to the presently available simulation and
theoretical results.

But, the concept of an average monomer, which is needed in the present
theory because, at variance with star polymers, there is no obvious
choice of a reference monomer, is rather artificial and in practice
conceals significant technical difficulties. Indeed, this concept has
a simple meaning only in the limit of very long chains, where all
monomers can be considered as equivalent. In Ref.~\cite{bollou02mm},
the effective interactions between central monomers and between end
monomers have been found to differ considerably, showing that strong
finite length effects should be expected on effective interactions
between average monomers. Thus, for a better control of these finite
size effects, an alternative approach seems desirable, and it has been
found that considering effective interactions between polymer c.m.'s
is appropriate in this respect \cite{olalanpel80mm,dauhal94mm,%
loubolhanmei00prl,bollouhanmei01jcp,bollou02mm}.

\section{Effective interaction between centers of mass}\label{vcc.sec}

As has been shown in recent Monte Carlo studies
\cite{loubolhanmei00prl,bollouhanmei01jcp,bollou02mm,krahanlou03pre},
an effective pair potential $v_{cc}(r)$ between the c.m.'s of polymer
chains may be extracted by an inversion procedure analogous to the one
described in the previous Section. To that purpose one considers a
simple fluid of density $\rho$, the pair correlations of which are
characterized by $h_{cc}(r)$, the c.m. correlation function of the
polymer system. The corresponding direct correlation function will be
denoted $c_{cc}(r)$.

\begin{figure}
\includegraphics{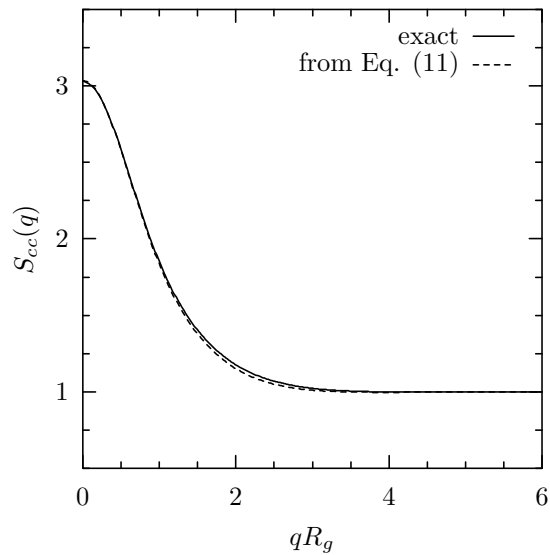}
\caption{\label{magic.fig} C.m. structure factor $S_{cc}(q)$ for
$L=500$ SAW polymers with monomer nearest-neighbor interaction
$\varepsilon$ on a simple cubic lattice, at $\beta\varepsilon=0.3$ and
$\rho R_g^3\simeq0.04$. The ``exact'' structure factor from MC
simulations is compared to the prediction of the approximation
(\ref{magic.eq}) using the ``exact'' monomer correlation function and
form factors.}
\end{figure}

The c.m. correlation function $h_{cc}(r)$ is easily accessible in MC
simulations for any polymer concentration
\cite{loubolhanmei00prl,bollouhanmei01jcp,bollou02mm,%
krahanlou03pre}, but cannot be calculated directly from the original
PRISM integral equation.  However an approximate relation between
$\widehat{h}_{mm}$ and $\widehat{h}_{cc}$ was established in
Ref.~\cite{krahanlou02el}, which reads
\begin{equation}\label{magic.eq}
\widehat{h}_{cc}(q)\simeq\left(\frac{\widehat{\Omega}_{cm}(q)}
{\widehat{\Omega}_{mm}(q)}\right)^2\widehat{h}_{mm}(q),
\end{equation}
where $\widehat{\Omega}_{mm}(q)$ is the previously discussed
intramolecular monomer-monomer form factor, and
$\widehat{\Omega}_{cm}(q)$ is the c.m.-monomer form factor [normalized
such that $\widehat{\Omega}_{cm}(q=0)=1$]. This relation was shown to
be quite accurate by comparing the resulting $\widehat{h}_{cc}(q)$, or
the corresponding c.m. structure factor $S_{cc}(q)=1+\rho
\widehat{h}_{cc}(q)$, to MC simulation data. To complete the results
of Ref.~\cite{krahanlou02el}, where only good solvent conditions were
explored, and to illustrate the quality of the predictions of
Eq.~(\ref{magic.eq}) for $S_{cc}(q)$, simulation results obtained for
a lattice model slightly below its $\theta$ temperature are reported
in Fig.~\ref{magic.fig}.

\begin{figure}
\includegraphics{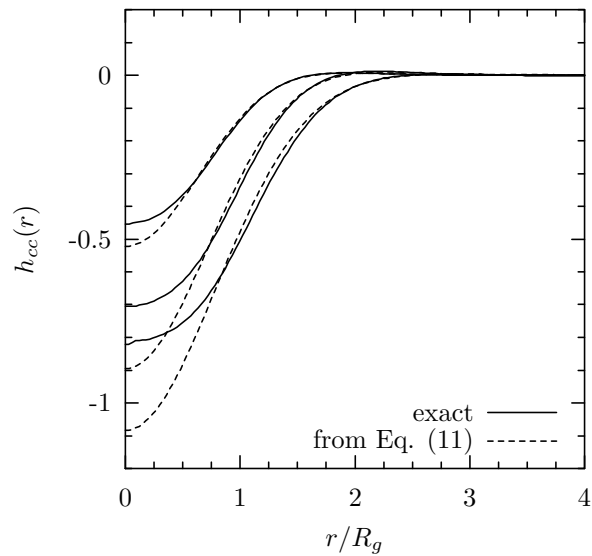}
\caption{\label{magicinr.fig} C.m. pair correlation functions
$h_{cc}(r)$ for $L=500$ SAW polymers on a simple cubic lattice. From
bottom to top, $\rho R_g^3\simeq 0.07,\ 0.27,\ 1.10$. The ``exact''
pair correlation functions are compared to the predictions of the
approximation (\ref{magic.eq}) using the ``exact'' monomer correlation
functions and form factors.}
\end{figure}

However, to use Eq.~(\ref{hnc.eq}), we explicitly need the pair
correlation function $h_{cc}(r)$, i.e. we have to formulate the
predictions of Eq.~(\ref{magic.eq}) in real space, and not in
reciprocal space. In Fig.~\ref{magicinr.fig}, we plot the simulation
results of Ref.~\cite{krahanlou02el} accordingly. The agreement
between the ``exact'' $h_{cc}(r)$ and its approximation computed from
Eq.~(\ref{magic.eq}), using the ``exact'' monomer correlation function
and form factors, is seen to be reasonably good, except at short
distances where the results of the approximation lie well below the
``exact'' simulation data, and can even become smaller than $-1$ at
low densities, which is unphysical. Consequences of this
underestimation of $h_{cc}(r)$ at full overlap, which is found under
the conditions of Fig.~\ref{magic.fig} as well and appears to be
systematic, are discussed below.

As mentioned in Sec.~\ref{prism.sec}, since there are no known
analytic expressions for the two form factors appearing in
Eq.~(\ref{magic.eq}) for interacting polymers, we are led to use the
expressions valid for Gaussian polymer coils in the scaling limit
\cite{koy80mc}, i.e. Eq.~(\ref{debye.eq}) and
\begin{equation}\label{omegacm.eq}
\widehat{\Omega}_{cm}(q)= \frac{\sqrt{\pi}}{x}e^{-x^2/12}
\text{erf} \left(\frac{x}{2}\right),\ x=qR_g,
\end{equation}
where $\text{erf}$ denotes the error function. In
Ref.~\cite{krahanlou02el}, it was shown that they yield qualitatively
correct results for the calculation of $\widehat{h}_{cc}(q)$, even in
the good solvent regime.

\begin{figure*}
\includegraphics{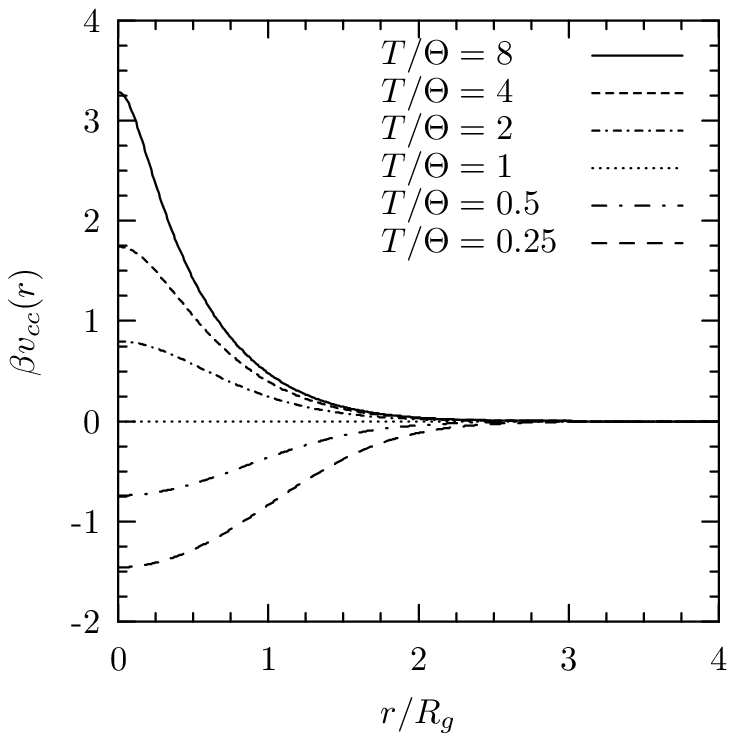}
\hspace{0.5cm}
\includegraphics{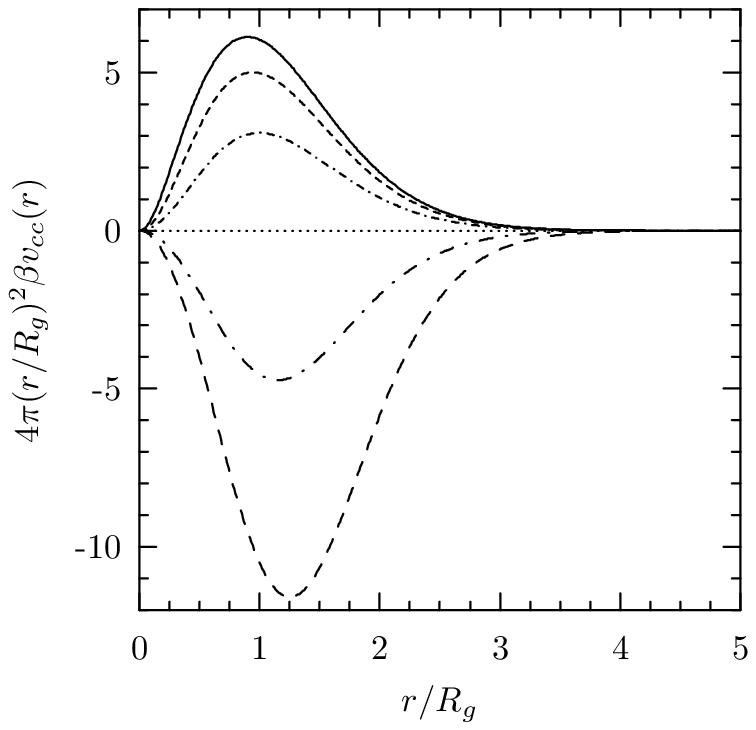}
\caption{\label{vcctemp.fig} Zero density effective pair potentials
between centers of mass (left-hand frame), for various values of the
temperature ratio $T/\Theta$; the right-hand frame shows the
potentials multiplied by $4\pi(r/R_g)^2$.}
\end{figure*}

The importance of using exact results is best seen when the high
density limit of $\widehat{h}_{cc}(q)$ for $T>\Theta$ is considered,
which follows from Eq.~(\ref{h.eq}):
\begin{equation}
\lim_{\rho\to\infty} \rho \widehat{h}_{cc}(q)=
-\frac{\widehat{\Omega}_{cm}(q)^{2}}{\widehat{\Omega}_{mm}(q)}.
\end{equation}
Indeed, while this ratio is always larger than $-1$ if the exact form
factors (\ref{debye.eq}) and (\ref{omegacm.eq}) are combined, it shows
a region at small $q$ where it is smaller than $-1$ if the
approximation (\ref{omegaapp.eq}) is used. In the latter case,
spurious singularities are thus encountered when Eq.~(\ref{oz.eq}) is
applied to compute $\widehat{c}_{cc}(q)$ at high densities.

In this high density regime, an expression for the potential of mean
force between the c.m.'s of polymers in the melt has been derived by
Guenza, on the basis of the same thread PRISM theory as the one used
here, in the case of excluded volume interactions only
\cite{gue02prl}. Both approaches share similarities, like the fact
that, to first order, the potential of mean force vanishes
proportionally to $1/\rho$. However, since many aspects of these
theories differ and since we are interested in the effective potential
and not the potential of mean force, a detailed comparison was not
attempted.

The outline of the present computation is thus the following. First,
the thread PRISM theory developed in Sec.~\ref{prism.sec} is used to
obtain $h_{mm}(r)$. Then $h_{cc}(r)$ is derived by applying
Eq.~(\ref{magic.eq}) with form factors (\ref{debye.eq}) and
(\ref{omegacm.eq}), and is eventually inverted using
Eqs.~(\ref{oz.eq}) and (\ref{hnc.eq}) (with indices $c$ instead of
$a$) to obtain $\beta v_{cc}(r)$.

We first consider the zero density limit.  When $\rho\equiv0$, the
effective pair potential is simply the potential of mean force, i.e.
$\beta v_{cc}(r)=-\ln [1+h_{cc}(r)]$. Combining Eqs.~(\ref{h.eq})
(with $\rho=0$) and (\ref{magic.eq}), one finds
\begin{equation}\label{meanfieldish.eq}
\beta v_{cc}(r)=-\ln [1+\widetilde{c} R_g^3
\Omega_{cm}*\Omega_{cm}(r)],
\end{equation}
where $*$ denotes a space convolution.  Results for $\beta v_{cc}(r)$
and $4\pi(r/R_g)^2\beta v_{cc}(r)$ are shown in Fig.~\ref{vcctemp.fig}
for several temperatures. Above $\Theta$, $\beta v_{cc}(r)$ has a
roughly Gaussian shape, with an amplitude which decreases with
temperature. The potential vanishes identically for $T=\Theta$ and
becomes purely attractive below, retaining a roughly Gaussian shape
with a negative amplitude increasing in absolute value when $T$ is
decreased. These trends are globally in agreement with the behavior
found in MC simulations \cite{dauhal94mm,krahanlou03pre}, but
quantitative differences are visible.

In the good solvent regime, for temperatures above approximately
$11\Theta$, it is impossible to compute $\beta v_{cc}(r)$. This is due
to the theory producing unphysical c.m. pair distribution functions,
which are negative in the vicinity of $r=0$. Moreover, approaching
this existence limit from below, the value of $\beta v_{cc}(r)$ at
full overlap is found to diverge, corresponding to the fact that
$h_{cc}(r=0)$ goes to $-1$. This behavior is at odds with all
available simulation data
\cite{olalanpel80mm,dauhal94mm,loubolhanmei00prl,bollouhanmei01jcp,%
bollou02mm,krahanlou03pre} and other theoretical approaches
\cite{grokhakho82mcrc,kruschbau89jpf}, showing that in good solvent
the amplitude of the potential of mean force is always finite and has
a maximum of about $2 k_{B}T$, a value reached for $T\simeq5\Theta$ in
the present theory. This deficiency, which manifests itself throughout
the high-$T$, low-$\rho$ domain, is an obvious consequence of the
underestimation of $h_{cc}(r)$ at full overlap, as illustrated in
Fig.~\ref{magicinr.fig}: While the exact $h_{cc}(r=0)$ saturates at a
value strictly greater than $-1$ in the limit of infinite temperature,
it is predicted by the present theory to go to $-1.1$ in the same
limit.  At this point it is worth insisting on the fact that this
problem appears to be associated with the use of Eq.~(\ref{magic.eq}),
and is totally unrelated to the use of Gaussian chain form factors to
model good solvent conditions.

Around and below $\Theta$, a more complex $r$ dependence of $\beta
v_{cc}(r)$ was found in the simulations of
Refs.~\cite{dauhal94mm,krahanlou03pre}, the potential globally growing
more and more negative, as found here, but retaining a repulsive
component at short distances, while it is found strictly monotonous in
the present calculation. These two types of variations were also found
in various theoretical approaches based on mean-field ideas
\cite{flokri50jcp,tan85jcp,czehal91mm,grokuz92mm,raoall96mm}, and in
fact it is not clear what the generic behavior of $\beta v_{cc}(r)$ is
in poor solvents. The available simulation data were indeed obtained
for rather short chains, for which, on the scale of $R_g$, the monomer
hard cores are not of negligible size. This could result in the
observed short range repulsive contribution to $\beta v_{cc}(r)$,
which would then be a finite size effect disappearing for longer
chains \cite{repulsion}.

\begin{figure*}
\includegraphics{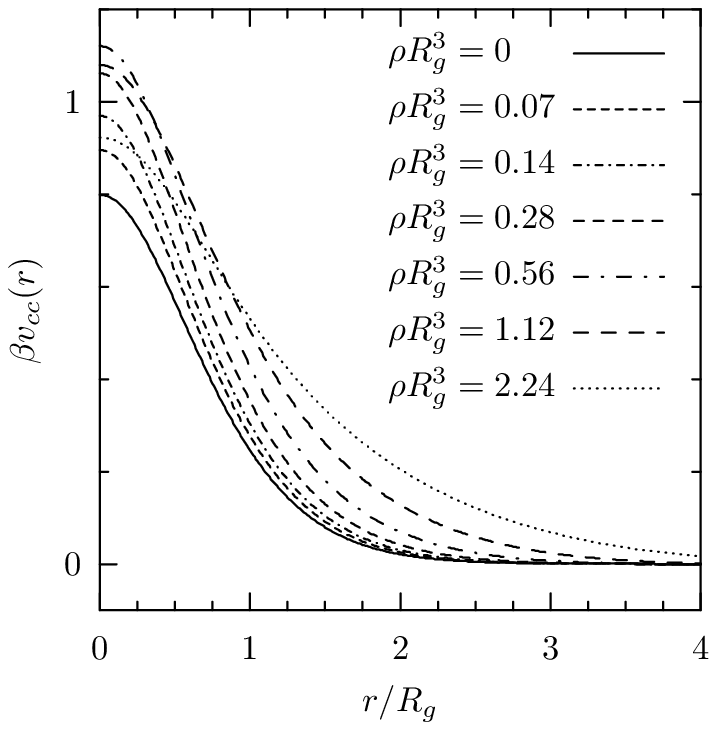}
\hspace{0.5cm}
\includegraphics{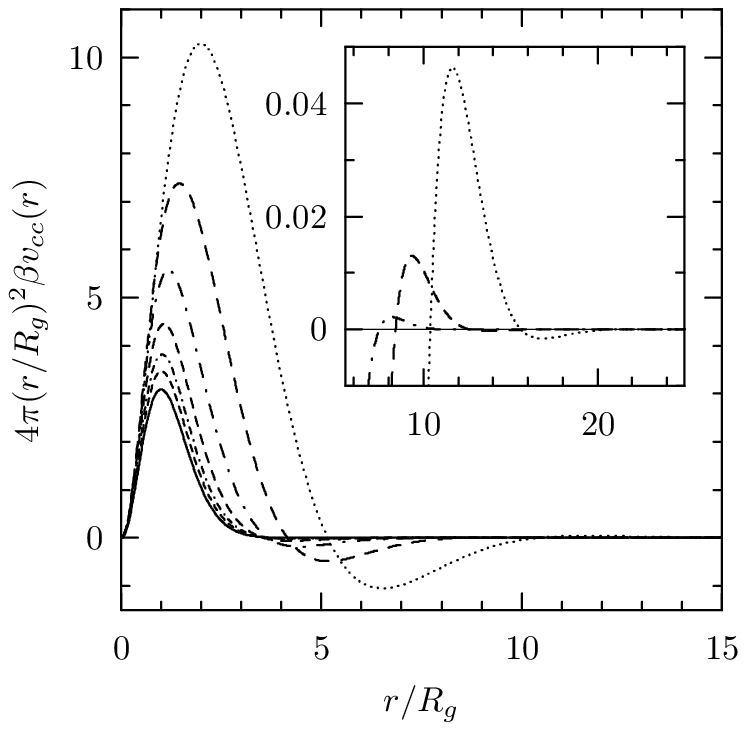}
\caption{\label{vcc.fig} Effective pair potentials between centers of
mass (left-hand frame), at temperature $T=2\Theta$, for various values
of the reduced density $\rho R_g^3$; the right-hand frame shows the
potentials multiplied by $4\pi(r/R_g)^2$, and the insert shows an
enlargement of this function for large $r/R_g$.}
\end{figure*}

By comparison with mean-field theories, difficulties can in fact be
anticipated with the present theory in the low temperature
domain. Indeed, in the zero density limit, following
Eq.~(\ref{meanfieldish.eq}), $\beta v_{cc}(r)$ is a simple function of
the overlap between the monomer distributions around the centers of
mass of two coils at distance $r$. However, mean-field theories built
exclusively on this two polymer overlap are known to take into account
only two-body effective monomer-monomer interactions. This is for
instance the case of the theory of Flory and Krigbaum
\cite{flokri50jcp,czehal91mm}, which can be recovered in a thread
PRISM version by linearizing Eq.~(\ref{meanfieldish.eq}). Such
approaches are valid in good solvent, since then pair interactions
actually dominate, but they fail in $\Theta$ and poor solvents, where
inclusion of higher order interactions is required. By analogy, this
raises questions on the applicability of the present theory in the
latter domain as well.

We now turn to the density dependence of the effective potentials
derived within the present theory.  Effective potentials obtained for
various densities at $T=2\Theta$ are plotted in
Fig.~\ref{vcc.fig}. These results are typical of good solvent
conditions. Starting from $\rho R_g^3=0$, where the potential is
purely repulsive and nearly Gaussian, $v_{cc}(r)$ first increases in
amplitude and in range with $\rho R_g^3$, then, at a density which
increases when the temperature approaches $\Theta$ from above, the
trend changes: the repulsion at full overlap decreases as $\rho R_g^3$
is further increased, while the range of the effective potential still
increases. In addition, with increasing $\rho R_g^3$, a damped
oscillatory tail, which is more clearly seen on plots of $4\pi
(r/R_g)^2 \beta v_{cc}(r)$, progressively develops when the distance
between the c.m.'s of two coils greatly exceeds $R_g$. All these
trends are consistent with the simulation data of
Refs.~\cite{loubolhanmei00prl,bollouhanmei01jcp,bollou02mm,krahanlou03pre}.
The only qualitative difference is in the tail of the potential,
which, within the resolution of the simulations, appears to be purely
attractive. The results of the present theory suggest that to unveil a
possible oscillatory behavior, simulations of much larger bulk systems
with very high statistical accuracy would be required to characterize
the potential beyond its first minimum.

\begin{figure*}
\includegraphics{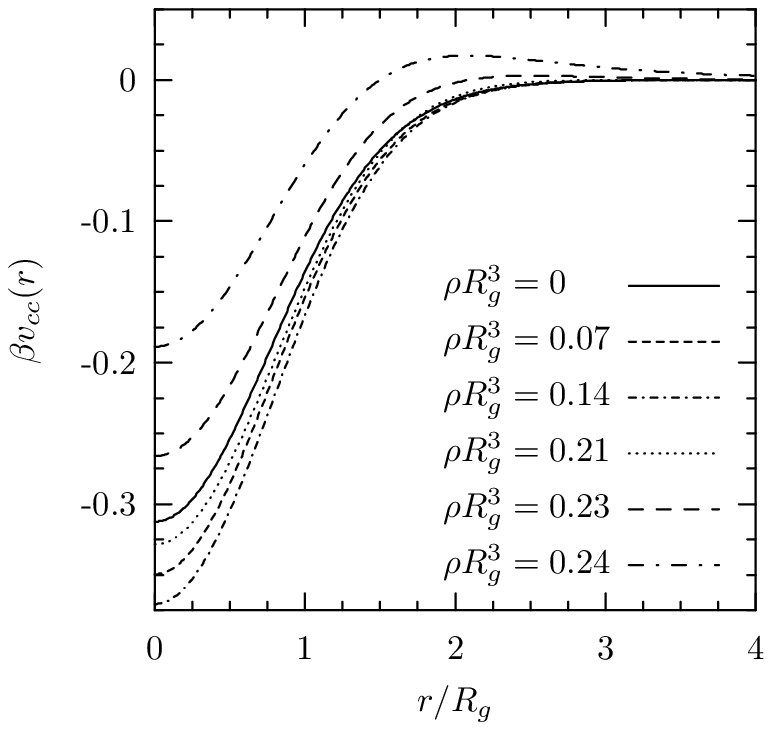}
\hspace{0.5cm}
\includegraphics{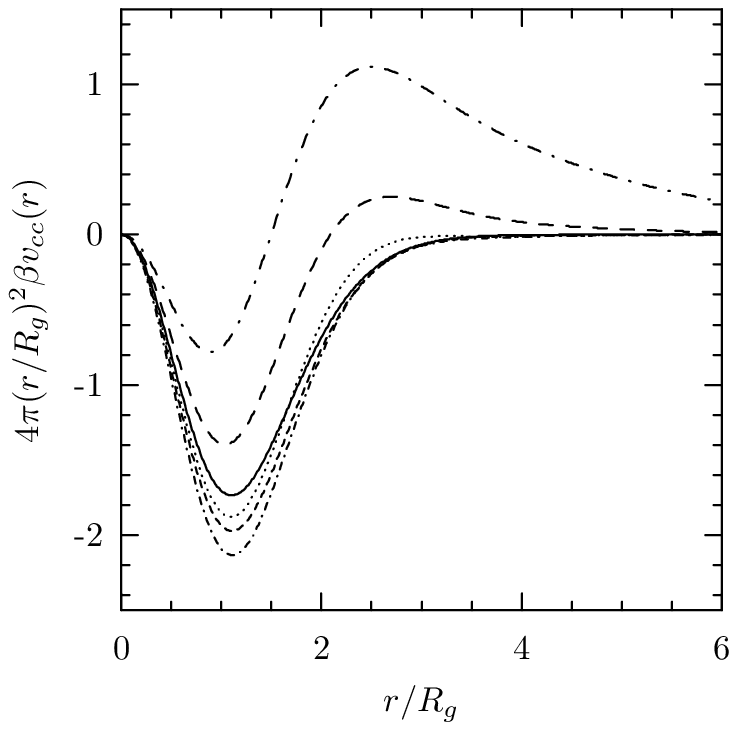}
\caption{\label{vcc2.fig} Effective pair potentials between centers of
mass (left-hand frame), at temperature $T=0.75\Theta$, for various
values of the reduced density $\rho R_g^3$; the right-hand frame shows
the potentials multiplied by $4\pi(r/R_g)^2$.}
\end{figure*}

Despite the various concerns discussed above about the applicability
of the present theory around and below the $\theta$ temperature (lack
of more-than-two-body effective monomer-monomer interactions,
incomplete characterization of the domain of existence of the low
density phase \cite{binodal}), we present the corresponding results
for completeness.  When $T=\Theta$, the effective c.m. potential is
found to vanish identically at all densities. Typical results for the
poor solvent domain, obtained for various densities at $T=0.75\Theta$,
are plotted in Fig.~\ref{vcc2.fig}.  Starting from $\rho R_g^3=0$,
where the potential is negative and nearly Gaussian, $v_{cc}(r)$ is
first found to become more negative with increasing $\rho R_g^3$,
while roughly retaining its shape. Then, upon approaching the
``spinodal'' instability line, at a density which increases when the
temperature approaches $\Theta$ from below, the value of the potential
at full overlap starts to grow with $\rho R_g^3$ and a repulsive tail
progressively develops.  These tendencies are not corroborated by the
presently available simulation data.

\section{Discussion and conclusion}

We have developed an integral equation approach to the computation of
effective interactions between polymers in solution. It is based on
the thread limit of the PRISM theory, which is used directly for the
calculation of effective interactions between average monomers, and in
association with an approximate expression for the c.m. pair
distribution function, derived within the PRISM formalism as well, for
the calculation of c.m.-c.m. effective potentials.

In the case of the effective interactions between average monomers,
the predictions of the theory for polymers with excluded volume
interactions only are in good qualitative agreement with the available
simulation data. Moreover, we have shown that the theory lends itself
to fully analytic calculations and compares rather favorably with
other analytic theories like those developed for star polymers.

The effective c.m.-c.m. interaction potentials derived from the
present theory exhibit many of the trends observed in MC simulations.
The agreement with the latter is very satisfactory at intermediate
temperature under good solvent conditions, i.e. $T>\Theta$. The theory
indeed reproduces non-trivial features like the non-monotonic
variation of the value of the potential at full overlap and the
development of a long range tail when the density increases.  This
agreement deteriorates at very high temperature, in particular for
$\beta=0$ (SAW model), and around and below the $\theta$-temperature.

Since the HNC closure used to compute the effective pair potential is
exact for all practical purposes, when applied to ``soft-core''
systems which exhibit essentially ``mean-field fluid'' behavior
\cite{lanlikwatlow00jpcm}, the observed discrepancies must be traced
back to the other approximations made in the theory, namely the
RMPY/HTA closure used to compute $h_{mm}(r)$ in the thread limit, the
use of form factors $\widehat{\Omega}_{mm}(q)$ and
$\widehat{\Omega}_{cm}(q)$ valid for ideal rather than interacting
chains, and Eq.~(\ref{magic.eq}) relating $\widehat{h}_{cc}(q)$ to
$\widehat{h}_{mm}(q)$. Simulation data tend to show that the use of
Eq.~(\ref{magic.eq}) is the main cause of the breakdown of the theory
at very high temperature, while there are indications that the present
use of the RMPY/HTA closure in the low temperature regime is not
justified. In the present state of the theory, the use of ideal chain
form factors is in fact not found to be a critical approximation.

In conclusion, we have proposed a first step towards an integral
equation theory of effective pair potentials between macromolecules in
solution, which could bypass costly simulations for the determination
of the c.m.  pair correlation function. The results appear very
encouraging, since, while only the simplest ingredients have been
combined in the present theory, and there is clearly room for
improvement, it is already found to be able to capture various
non-trivial behaviors observed in recent simulations. It appears thus
that integral equation theories could be in the future a valuable tool
for the development of coarse-grained descriptions of complex
polymeric systems.

\acknowledgments 

The authors thank A.A. Louis for useful discussions and
B.R. acknowledges financial support from the \'Ecole Normale
Sup\'erieure (Paris) and Schlumberger Cambridge Research.

\appendix
\section{Thread PRISM calculation}\label{thread.app}

In this Appendix, we present our calculation of the monomer pair
correlation function based on PRISM theory in the thread limit. It is
inspired by the work of Chatterjee and Schweizer
\cite{chasch98jcp,chasch98mm} (see also Ref.~\cite{cha93pre} for a
related approach), a special asymptotic regime of which is obtained.
We start with Eq.~(\ref{prismozthread.eq}), involving quantities which
all remain finite in the thread limit as defined in
Sec.~\ref{prism.sec}. The case of excluded volume interactions only is
considered first, and then used as a reference system to include
attractive interactions between the monomers in the closure relations.

\subsection{Reference system: Excluded volume interactions}

Considering first the excluded volume interaction only, a
Percus-Yevick (PY)-like closure is used, in which
$C_{mm}(r)=\widetilde{c}_{0} R_g^3 \delta(\mathbf{r})$ ($R_g^3$ is
introduced here to make $\widetilde{c}_{0}$ dimensionless), or
equivalently in Fourier space,
\begin{equation}
\widehat{C}_{mm}(q) = \tilde{c}_{0} R_g^3.
\end{equation}
This singular $C_{mm}(r)$ is required to account for the effect of
excluded volume on intermolecular correlations in the limit of
vanishing size of the hard core of the monomers in the thread limit.

The PRISM-OZ relation (\ref{prismozthread.eq}) then leads to
Eq.~(\ref{h.eq}), with $\widetilde{c}=\widetilde{c}_{0}$, and the
value of $\widetilde{c}_{0}$ is determined by enforcing the hard core
condition $h_{mm}(r=0)=-1$.

In general, all the calculations have to be performed
numerically. This is the case in the present work when
$\widehat{\Omega}_{mm}(q)$ is chosen to be the Debye function
(\ref{debye.eq}). However, fully analytic results can be obtained if
the usual Pad\'e approximant of this function,
Eq.~(\ref{omegaapp.eq}), is used. Then, one finds
\begin{equation}\label{czero.eq}
\widetilde{c}_{0}=-2\pi\sqrt{2}-2\pi^2\rho R_{g}^{3},
\end{equation}
and
\begin{equation}\label{hzero.eq}
h_{mm}(r) =\frac{1}{2\pi\rho R_{g}^{3}}\times
\frac{e^{-r/\xi_{\rho}}-e^{-r/\xi_{c}}}{r/R_g},
\end{equation}
where $\xi_{c}=R_{g}/\sqrt{2}$ and $\xi_{\rho}=R_g/(\sqrt2 + 2 \pi
\rho R_{g}^{3})$.

\subsection{Including an attractive tail}

We now introduce the attractive interaction between the monomers,
which, for convenience, is taken of the Yukawa form (\ref{yukawa.eq}).
In Ref.~\cite{chasch98jcp}, various closures of the PRISM-OZ equation
have been studied in the thread limit, for chains with monomers
interacting through a hard core and an attractive tail. Here, we
consider the reference molecular Percus-Yevick/high temperature
approximation (RMPY/HTA) \cite{schyet93jcp,chasch98mm,chasch98jcp},
which has been found to be the most satisfactory in this earlier
work. With the present notations, it reads
\begin{equation}
C_{mm}(r) = C_{\text{ref}}(r)-\beta V(r) g_{\text{ref}}(r),
\end{equation}
where a renormalized attractive potential $V(r)=L^{2}v(r)$ was
introduced. $C_{\text{ref}}(r)$ and $g_{\text{ref}}(r)$ are respectively
the renormalized monomer direct correlation function and the monomer
pair distribution function of the reference system, the calculation of
which is explained above.

As mentioned in Sec.~\ref{prism.sec}, on physical grounds, we
supplement the previously applied limiting process with the limit of a
vanishing range of the attractive potential, namely $a\to0$, keeping
the ratio $\Delta=a/\sigma$ constant.  To take this supplementary
limit in the RMPY/HTA closure, we need to evaluate
$\lim_{a\rightarrow0}\beta V(r) g_{\text{ref}}(r)$. We assume that
$g_{\text{ref}}(r)$ satisfies the hard-core condition,
$g_{\text{ref}}(0)=0$, and has a finite slope at the origin. This
hypothesis holds both for the analytic result, Eq.~(\ref{hzero.eq}),
and when the exact Debye form factor is used instead of its Pad\'e
approximant (see below). Then we can write the Taylor expansion
\begin{equation}
g_{\text{ref}}(r) = g_{\text{ref}}'(0) r + o(r).
\end{equation}
Therefore
\begin{equation} \label{betagv.eq}
\begin{split}
\beta V(r) g_{\text{ref}}(r) & = - L^{2} \beta \epsilon a e^{-r/a}
g_{\text{ref}}'(0) + o(1) \\ & = -36 \beta \epsilon \Delta^{4}R_{g}^{4}
\frac{e^{-r/a}}{a^{3}} g_{\text{ref}}'(0) + o(1),
\end{split}
\end{equation}
where $R_g^2=L\sigma^2/6$ has been used.  Now, taking the limit
$a\rightarrow0$, we find
\begin{equation}
\lim_{a\rightarrow0}\beta V(r) g_{\text{ref}}(r)= -288 \pi \beta
\epsilon\Delta^{4}R_{g}^{4} g_{\text{ref}}'(0) \delta(\mathbf{r}),
\end{equation}
since the contribution of the $o(1)$ in (\ref{betagv.eq}) goes to zero.
The RMPY/HTA closure relation hence reduces to
\begin{equation}\label{c2.eq}
C_{mm}(r) = \widetilde{c}_0 \left(1-\frac{\Theta}{T}\right)
R_g^3 \delta(\mathbf{r}),
\end{equation}
with the temperature $\Theta$ given by
\begin{equation}
\Theta = - 288 \pi \Delta^{4} \frac{g_{\text{ref}}'(0)
R_g}{\widetilde{c}_0}\frac{\epsilon}{k_{B}}
\end{equation}
[remember that $\widetilde{c}_0<0$, as seen in Eq.~(\ref{czero.eq})].
The latter expression can be greatly simplified, using the fact that,
for large $q$, $\widehat{h}_{\text{ref}}(q)$ decays like
$4\widetilde{c}_{0}/( q^{4}R_g)$; this allows to show that
\begin{equation}
g_{\text{ref}}'(0) = -\frac{\widetilde{c}_0}{2 \pi R_{g}},
\end{equation}
which yields
\begin{equation}\label{theta.eq}
\Theta = 144 \Delta^{4} \frac{\epsilon}{k_{B}}.
\end{equation}
Thus, we find that $\Theta$ is the same for both choices of the
polymer form factors, Eq.~(\ref{debye.eq}) or (\ref{omegaapp.eq}), and
is independent of the polymer density.  Fourier transforming
$C_{mm}(r)$, defining $\widetilde{c}$ as in Eq.~(\ref{c.eq}), and
injecting the result in Eq.~(\ref{prismozthread.eq}), one obtains
Eq.~(\ref{h.eq}).

As anticipated by its notation, $\Theta$ is in fact the $\theta$
temperature of the present polymer model. This can be first seen from
Eq.~(\ref{c2.eq}), since one finds that $C_{mm}(r)$, often interpreted
as a renormalized two-body monomer-monomer interaction, vanishes for
$T=\Theta$. More evidently, this can be shown by considering the
pressure of the polymer fluid. Following the ``compressibility route''
\cite{simpleliquids}, the pressure is indeed easily calculated to be
\begin{equation}\label{eost.eq}
\begin{split}
\beta P &= \int_{0}^{\rho}[1-\rho'\widehat{C}_{mm}(q=0,\rho')]d\rho'
\\ &= \rho - \left(1-\frac{\Theta}{T}\right) R_{g}^{3} \int_{0}^{\rho} 
\rho' \widetilde{c}_0(\rho') d\rho',
\end{split}
\end{equation}
and one sees that, for $T=\Theta$, the pressure is that of a perfect
gas, meaning in particular that its second virial coefficient
vanishes, which is the usual definition of the $\theta$ temperature.

Here again, a fully analytic expression for $h_{mm}(r)$ can be
obtained if the Pad\'e approximant of the chain form factor is
used. It retains the form (\ref{hzero.eq}), with $\xi_c$ unchanged and
$\xi_\rho=R_g/\sqrt{2(1-\rho \widetilde{c}R_g^3)}$.

\subsection{Relation to the theory of Chatterjee and Schweizer}

We show here that the present calculation is related to an asymptotic
regime of the theory developed in Refs.~\cite{chasch98jcp,chasch98mm},
in which the range of the attractive tail was kept finite. To this
purpose, the calculation of the pressure is considered.

In Refs.~\cite{chasch98jcp,chasch98mm}, where the approximate chain
form factor (\ref{omegaapp.eq}) was used, the pressure was found to be
given by (see Eq.~(30) of Ref.~\cite{chasch98jcp})
\begin{multline}\label{pressure.eq}
\beta P = \rho + \pi \sqrt{2} R_g^3 \rho^2 +\frac23 \pi^2 R_g^6 \rho^3
- \beta\varepsilon \times \\ \left( 12 \pi \sqrt{6} \Delta^3 R_g^3
\sqrt{L} \rho^2 - \frac{12 \Delta^2 L \rho}{1+(a\sqrt{2}/R_g)} +
\right.\\ \left.\frac{\sqrt{6} \Delta L \sqrt{L}}{\pi R_g^3}\ln
\left[1 + \frac{2\pi\sqrt{6} \Delta R_g^3 \rho}{\sqrt{L} [1+(a
\sqrt{2}/R_g)]} \right]\right),
\end{multline}
where the notations of the present paper are used and $\sigma$ and
$\rho_m$ have been reexpressed as functions of $L$, $R_g$, and $\rho$,
using the relations $R_g^2=L \sigma^2/6$ and $\rho_m=\rho L$. $a$ can
be handled in a similar way, using the fact that $R_g^2= L a^2 /(6
\Delta^2)$, and all the variables which vanish in the thread limit as
defined in Sec.~\ref{prism.sec} are then eliminated. The only
remaining singular quantity is $L$, which should go to
infinity. Taking this limit in Eq.~(\ref{pressure.eq}), after a large
$L$ expansion of the logarithm and denominators is performed, leads to
\begin{equation}
\beta P= \rho + \pi \sqrt{2} R_g^3 \left(1-\frac{\Theta}{T}\right)
\rho^2 +\frac23 \pi^2 R_g^6 \left(1-\frac{\Theta}{T}\right) \rho^3,
\end{equation}
which is precisely the result obtained by combining
Eqs.~(\ref{eost.eq}) and (\ref{czero.eq}). This indicates that the
present theory corresponds to an $a\to0$ limit of the theory of
Chatterjee and Schweizer.

\subsection{Limitations of the model}

The present polymer theory shows some limitations which are reviewed
here.

First, in the presence of attractive interactions, the hard core
condition $h_{mm}(r=0)=-1$ is no longer enforced
\cite{chasch98jcp}. For instance, one finds for the analytic solution
\begin{equation}
h_{mm}(r=0) =\frac{1-\sqrt{1-\rho\widetilde{c}R_g^3}}{\pi\sqrt{2}\rho
R_{g}^{3}},
\end{equation}
which is strictly larger than $-1$ for all densities and finite
temperatures. Such a feature is very common in this kind of
thermodynamic perturbation type approximations, where the direct
correlation function is completely specified for all values of $r$. We
expect its impact to be modest on the outcome of the proposed
coarse-graining procedure, since we will be interested in a
description on a mesoscopic level, while this inaccuracy manifests
itself on the microscopic, monomer level.

Secondly, below the $\theta$ temperature, there is a domain where the
integral equations exhibit a singularity, corresponding to the
vanishing of the denominator of Eq.~(\ref{h.eq}).  This is in fact the
remnant, when $L$ goes to infinity while taking the thread limit used
in this work, of the finite $L$ spinodal instability found by
Chatterjee and Schweizer \cite{chasch98jcp}, with a critical
temperature equal to $\Theta$ and an infinite critical value of $\rho
R_g^3$.  The boundary of this low-temperature, high-density domain can
be explicitly computed when the approximate expression of the form
factor is used, leading to
\begin{equation}
\rho R_g^3=\frac{1}{\pi \sqrt2}
\left[\sqrt\frac{\Theta}{\Theta-T}-1\right],\ T<\Theta.
\end{equation}
This means that, below the $\theta$ temperature, where polymer
solutions are known to phase separate between polymer-rich and
solvent-rich fluids, the proposed theory will only allow to
investigate the latter phase.

\section{Analytic effective pair potential between average monomers}
\label{anavaa}
A fully analytic expression for $\beta v_{aa}(r)$ can be obtained if
the approximate expression (\ref{omegaapp.eq}) for the chain form
factor is used. Indeed, the expression for $h_{aa}(r)\equiv h_{mm}(r)$
is known; solving Eq.~(\ref{oz.eq}) for $h_{aa}(q)$ and identifying
the result with Eq.~(\ref{h.eq}), one finds
\begin{equation}
\widehat{c}_{aa}(q) = \frac{\widetilde{c} R_g^3
\widehat{\Omega}_{mm}(q)^{2}}{1-
\rho\widetilde{c}R_g^3\widehat{\Omega}_{mm}(q)+
\rho\widetilde{c}R_g^3\widehat{\Omega}_{mm}(q)^2}.
\end{equation}
Replacing $\widehat{\Omega}_{mm}(q)$ by its expression
(\ref{omegaapp.eq}), $\widehat{c}_{aa}(q)$, and subsequently
$c_{aa}(r)$, are easily obtained. Two regimes are found, depending on
the sign of $\widetilde{c}$.

If $\widetilde{c}<0$, i.e. $T>\Theta$, one finds that $c_{aa}(r)$ is
the difference between two Yukawa terms,
\begin{equation}
c_{aa}(r) = \frac{\widetilde{c}}{2\pi
\sqrt{\rho\widetilde{c}R_g^3 (\rho\widetilde{c}R_g^3-4)}}
\times \frac{e^{-r/\xi_{+}}-e^{-r/\xi_{-}}}{r/R_g},
\end{equation}
where
\begin{equation}
\xi_\pm=\frac{R_g}{\sqrt{2-\rho\widetilde{c}R_g^3\mp
\sqrt{\rho\widetilde{c}R_g^3 (\rho\widetilde{c}R_g^3-4)}}}.
\end{equation}
If $\widetilde{c}>0$, i.e. $T<\Theta$, $c_{aa}(r)$ shows damped
oscillations,
\begin{equation}
c_{aa}(r) = -\frac{\widetilde{c}}{\pi
\sqrt{\rho\widetilde{c}R_g^3 (4-\rho\widetilde{c}R_g^3)}}
\times \frac{e^{-r/\xi_e}}{r/R_g}\sin (r/\xi_s),
\end{equation}
where
\begin{equation}
\xi_e=R_g\sqrt\frac{2}{4-\rho\widetilde{c}R_g^3},\
\xi_s=R_g\sqrt\frac{2}{\rho\widetilde{c}R_g^3}.
\end{equation}

Combining these results with Eqs.~(\ref{hzero.eq}) and (\ref{hnc.eq})
provides the desired analytic expression for $\beta v_{aa}(r)$.


\begin{thebibliography}{10}
\bibitem{order} In this practical calculation, the order of the
operations of monomer averaging and effective potential calculation
has been inverted, compared to the description of the process given
before. In the limit of very long chains where end effects are
negligible, this should be harmless. However, there is no reason to
think that in general these two operations commute and this could lead
to subtle effects, for short chains for instance.
\bibitem{zeroexp} For $\rho\equiv 0$, $\beta v_{aa}(r)=-\ln[1-
e^{-r/\xi_c}]$, and thus decays only exponentially for large
$r$. However, for all cases of interest, the polymer density is finite
and this marginal behavior is of little relevance.
\bibitem{binodal} The impossibility to properly identify the domain of
existence of the low density phase below $\Theta$ is a consequence of
the version of the thread PRISM theory considered in this work, which
implements the limits $L\to\infty$, $d$, $\sigma$, and $a\to0$ in a
strict sense. As can be seen in Appendix~\ref{thread.app} in the case
of the spinodal line, this rejects the liquid branch to infinite
polymer density and thus prevents a meaningful investigation of
liquid-gas coexistence. This was not the case in
Ref.~\cite{chasch98jcp}, where complete phase diagrams, including both
spinodals and binodals, could be obtained.
\bibitem{repulsion} Upon inversion of the low density simulation data
of Fig.~\ref{magic.fig} for chains of length $L=500$, the resulting
effective c.m.-c.m. pair potential does not show a repulsive short
range component. This tends to support the possibility that this
repulsion is a short chain artifact, but a more systematic study in
temperature and density would be necessary to reach a final
conclusion.
\end{thebibliography}
\end{document}